\documentclass[aps,prl,reprint,superscriptaddress,showpacs]{revtex4-1} 
\usepackage{graphicx}
\usepackage{dcolumn}
\usepackage{mathtools}
\usepackage{color}
\usepackage{braket}
\usepackage{hyperref}
\usepackage{soul}

\begin{document}

\title{End-to-end correlated subgap states in hybrid nanowires}

\author{G.~L.~R.~Anselmetti}
\affiliation{Center for Quantum Devices, Niels Bohr Institute, University of Copenhagen, and Microsoft Quantum - Copenhagen, Universitetsparken 5, 2100 Copenhagen, Denmark}
\author{E.~A.~Martinez}
\affiliation{Center for Quantum Devices, Niels Bohr Institute, University of Copenhagen, and Microsoft Quantum - Copenhagen, Universitetsparken 5, 2100 Copenhagen, Denmark}
\author{G.~C.~M\'enard}
\affiliation{Center for Quantum Devices, Niels Bohr Institute, University of Copenhagen, and Microsoft Quantum - Copenhagen, Universitetsparken 5, 2100 Copenhagen, Denmark}
\author{D.~Puglia}
\affiliation{Center for Quantum Devices, Niels Bohr Institute, University of Copenhagen, and Microsoft Quantum - Copenhagen, Universitetsparken 5, 2100 Copenhagen, Denmark}
\author{F.~K.~Malinowski}
\affiliation{Center for Quantum Devices, Niels Bohr Institute, University of Copenhagen, and Microsoft Quantum - Copenhagen, Universitetsparken 5, 2100 Copenhagen, Denmark}
\author{J.S.~Lee}
\affiliation{California NanoSystems Institute, University of California, Santa Barbara, California 93106, USA}
\author{S.~Choi}
\affiliation{California NanoSystems Institute, University of California, Santa Barbara, California 93106, USA}
\author{M.~Pendharkar}
\affiliation{Department of Electrical Engineering, University of California, Santa Barbara, California 93106, USA}
\author{C.~J.~Palmstr\o{}m}
\affiliation{California NanoSystems Institute, University of California, Santa Barbara, California 93106, USA}
\affiliation{Department of Electrical Engineering, University of California, Santa Barbara, California 93106, USA}
\affiliation{Materials Department, University of California, Santa Barbara, California 93106, USA}
\author{C.~M.~Marcus}
\affiliation{Center for Quantum Devices, Niels Bohr Institute, University of Copenhagen, and Microsoft Quantum - Copenhagen, Universitetsparken 5, 2100 Copenhagen, Denmark}
\author{L.~Casparis}
\email{Equal contribution, lucas.casparis@microsoft.com}
\affiliation{Center for Quantum Devices, Niels Bohr Institute, University of Copenhagen, and Microsoft Quantum - Copenhagen, Universitetsparken 5, 2100 Copenhagen, Denmark}
\author{A.~P.~Higginbotham}
\email{Equal contribution, andrew.higginbotham@ist.ac.at}
\affiliation{Center for Quantum Devices, Niels Bohr Institute, University of Copenhagen, and Microsoft Quantum - Copenhagen, Universitetsparken 5, 2100 Copenhagen, Denmark}
\affiliation{Institute of Science and Technology Austria, 3400 Klosterneuburg, Austria}

\date{\today}

\begin{abstract}
End-to-end correlated bound states are investigated in superconductor-semiconductor hybrid nanowires at zero magnetic field.
Peaks in subgap conductance are independently identified from each wire end, and a cross-correlation function is computed that counts end-to-end coincidences, averaging over thousands of subgap features.
Strong correlations in a short, $300~\mathrm{nm}$ device are reduced by a factor of four in a long, $900~\mathrm{nm}$ device.
In addition, subgap conductance distributions are investigated, and correlations between the left and right distributions are identified based on their mutual information.
\end{abstract}

\maketitle
Single Majorana bound states emerge at each end of a one-dimensional topological superconductor \cite{kitaev_unpaired_2001}, and pairs of Majorana bound states have been proposed to nonlocally encode quantum information \cite{kitaev_fault-tolerant_2003,nayak_non-abelian_2008}.
Following the theoretical suggestion that hybrid superconductor-semiconductor nanowires can possess a topological phase \cite{lutchyn_majorana_2010,oreg_helical_2010}, bound states within the superconducting gap (subgap states), have been extensively studied using the tunneling-conductance from a single wire end, and the results are broadly consistent with Majorana modes \cite{mourik_signatures_2012,das_zero-bias_2012,churchill_superconductor-nanowire_2013,deng_majorana_2016,zhang_quantized_2018}.
It has also been discovered, both in experiment \cite{lee_spin-resolved_2013,deng_nonlocality_2018} and theory \cite{kells_near-zero-energy_2012,prada_transport_2012,cayao_sns_2015,san-jose_majorana_2016,chun-xiao_andreev_2017,vuik_reproducing_2018,reeg_zero-energy_2018}, that localized non-topological or quasi-Majorana bound states can mimic many signatures of well-separated Majorana bound states.
Further, quantum-dot experiments can give information on the spatial extent of subgap states \cite{albrecht_exponential_2016,penaranda_quantifying_2018,deng_nonlocality_2018,whiticar_interferometry_2019}.
Probing both ends of the Majorana wire has been proposed to distinguish local states from Majorana modes by revealing end-to-end correlations between Majorana pairs, and bulk signatures of the topological transition \cite{dassarma_splitting_2012,fregoso_electrical_2013,stanescu_nonlocality_2014,rosdahl_andreev_2018,reeg_zero-energy_2018,lai_presence_2019}.

Independent tunneling spectroscopy of both wire ends  -- which is fundamentally required to assess the presence of end-to-end correlations -- has so far been challenging due to an inability to non-invasively ground the wire bulk.
Top-down lithography can interfere with the proximity effect \cite{krogstrup_epitaxy_2015,chang_hard_2015}, and semiconductor T-junctions have a poorly understood effect on the topological phase.
The recently-demonstrated selective area growth (SAG) of superconductor-semiconductor heterostructures offers an appealing solution to this problem, as it allows the lithographic processing on epitaxial contacts to be isolated from the delicate superconductor-semiconductor interface \cite{krizek_field_2018,vaitiekenas_selective-area-grown_2018}.
More broadly, SAG constitutes a potential platform for multi-terminal superconductor-semiconductor devices such as qubits \cite{larsen_semiconductor-nanowire-based_2015,casparis_superconducting_2018,plugge_majorana_2017,karzig_scalable_2017}, Cooper-pair splitters \cite{hofstetter_cooper_2009}, and multi-terminal Josephson junctions \cite{vanHeck_single_2014,strambini_omega-squipt_2016,meyer_nontrivial_2017,xie_topological_2017,pankratova_multi-terminal_2018}.

\begin{figure}[b]
    \centering
    \includegraphics[width=0.48\textwidth]{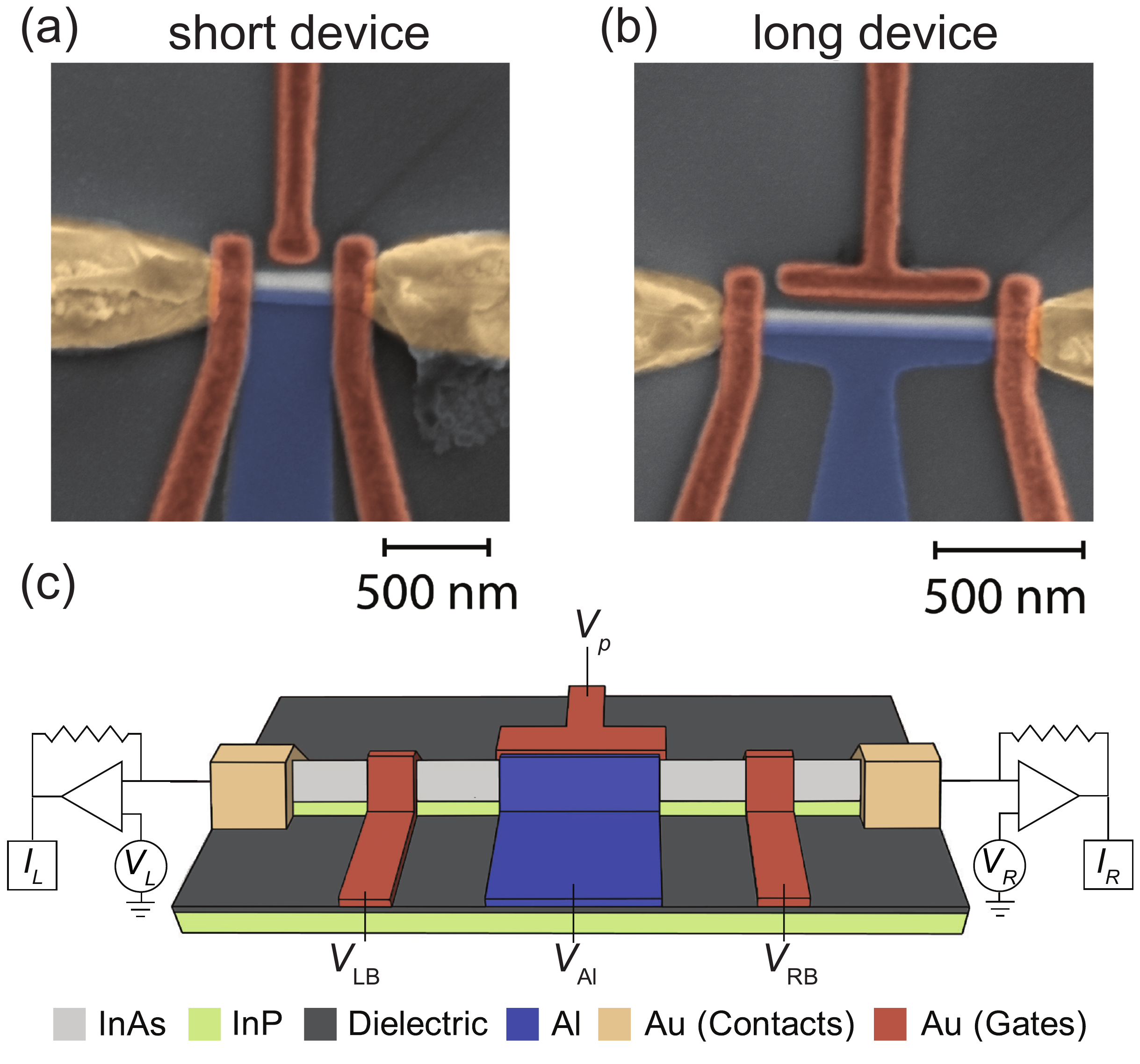}
    \caption{(a) Scanning electron micrograph of the short device with 300~nm superconducting segment.
    (b) Scanning electron micrograph of the long device with 900~nm superconducting segment.
    (c) Conceptual sketch of the devices showing the InP substrate, SiOx mask, Al lead (blue), Ti/Au Ohmic contacts (yellow), and Ti/Au electrostatic gates (red) separated by an HfO layer. Left bias voltage, $V_{L}$, left measured current, $I_L$, right bias voltage, $V_{R}$, right measured current, $I_R$, aluminum voltage, $V_\mathrm{Al}$, left-barrier voltage, $V_\mathrm{LB}$, right barrier voltage, $V_\mathrm{RB}$, and plunger voltage, $V_{P}$, labeled.
    }
    \label{fig1}
\end{figure}

In this Letter, we demonstrate that many of the challenges associated with studying end-to-end correlations in Majorana nanowires have been overcome.
We identify subgap features at both ends of three-terminal SAG devices at zero magnetic field, and quantify the correlations by averaging over thousands of subgap features.
Varying the length of the superconductor-semiconductor hybrid segment, we find that correlations are reduced by a factor of four in a long, $900~\mathrm{nm}$ nanowire, as compared to a short, $300\ \mathrm{nm}$ nanowire.
We also study the mutual information between the left and right subgap conductance distributions, finding subtle signatures of conductance correlations which are especially pronounced for a family of stable, zero-energy features.
This work demonstrates an experimental protocol for quantifying end-to-end correlations and determining the associated length scales.
Application of these results at non-zero magnetic field will enable searches for correlated Majorana pairs with statistical significance.

The platform for observing end-to-end correlations is Al-InAs SAG, grown by chemical beam epitaxy (CBE) on an InP substrate with a graded InAsP buffer layer~\cite{lee_selective-area_2018}, with epitaxial Al deposited \textit{in situ} immediately after semiconductor growth.
Following growth, Al (blue Fig.~1) was selectively removed from the substrate and the wire ends by a wet etch, and Ti/Au Ohmic contacts [yellow Fig.~1] were deposited on both ends of the wire.
The device was then covered by a global HfOx dielectric before electrostatic Ti/Au gates [red Fig.~1] are deposited.
Three devices were cofabricated, nominally differing only in the size of the superconducting region.
The short [$300\ \mathrm{nm}$, Fig.~1(a)] and long [$900\ \mathrm{nm}$, Fig.~1(b)] devices were studied in detail. An intermediate-length device did not show evidence of superconductivity and was damaged due to electrostatic discharge; it was not studied further.
All measurements were performed in a dilution refrigerator with base electron temperature $<100~\mathrm{mK}$.
To focus solely on the identification of end-to-end correlations without reference to the topological regime, the magnetic field is fixed to $B=0$; finite-field effects are explored in a partner paper \cite{menard_conductance-matrix_2019}.

Both devices have three terminals for electrical measurement [Fig.~1(c)].
Two normal-conducting contacts were used for tunneling spectroscopy, and were equipped with Basel SP983 I-to-V converters ($<~100~\mathrm{\Omega}$ input impedance).
The left normal contact sources a voltage $V_{L}$ and measures a current $I_L$.
Likewise, the right normal contact sources a voltage $V_{R}$ and measure a current $I_R$. 
The third terminal is formed from a selective etch of the epitaxial Al, and is set to a voltage $V_\mathrm{Al}$.
Because the third terminal is fabricated from a subtractive process after growth, it can be formed without disrupting the fragile Al-InAs interface.
Electrostatic gates, $V_\mathrm{LB}$ and $V_\mathrm{RB}$, tune the coupling to the left and right leads, and a plunger gate, $V_{P}$, tunes the electrochemical potential within the wire.
This setup allows the left conductance, $g_{L} = \delta I_L / \delta (V_{L} - V_\mathrm{Al})$, and the right conductance, $g_{R} = \delta I_R / \delta( V_{R} - V_\mathrm{Al} )$ to be independently measured using lock-in techniques.
The barriers are adjusted such that the above-gap conductance, averaged over gate voltage, is $< 0.15~\mathrm{e^2/h}$.

\begin{figure}
    \centering
    \includegraphics[width=0.48\textwidth]{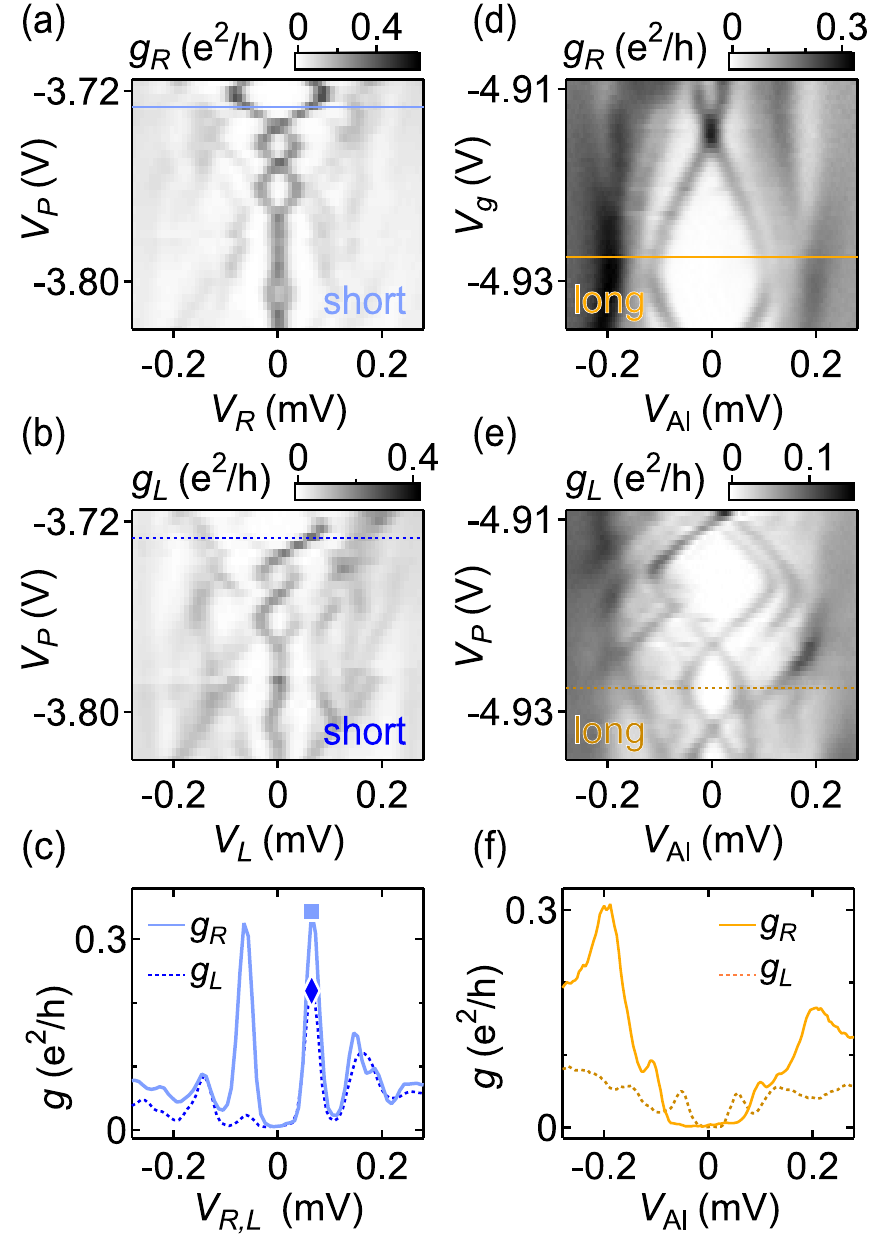}
    \caption{(a) Measured right conductance, $g_{R}$, versus plunger, $V_p$, and right bias, $V_{R}$.
    (b) Measured left conductance, $g_{L}$, versus plunger, $V_p$, and left bias, $V_{L}$.
    (c) Left and right conductance measured at fixed plunger voltage indicated by horizontal lines in 
    Fig.~1(a,b). $V_{R,L}$ denotes $V_{L}$ for $g_{L}$, and $V_{R}$ for $g_{R}$. Data are from the short device.
    (d) Measured right conductance, $g_{R}$, versus plunger, $V_p$, and middle lead bias, $V_{Al}$.
    (e) Measured right conductance, $g_{R}$, versus plunger, $V_p$, and middle lead bias, $V_{Al}$.
    (f) Left and right conductance measured at fixed plunger voltage indicated by horizontal lines in 
    Fig.~1(d,e). $V_{R,L}$ denotes $V_{L}$ for $g_{L}$, and $V_{R}$ for $g_{R}$. Data are from the long device.
    }
    \label{fig2}
\end{figure}

To study subgap structures in the short device, the right and left conductance, $g_{R}$ and $g_{L}$, were measured as a function of bias and plunger gate with $V_\mathrm{Al}=0$ fixed.
The data were obtained by sweeping $V_{R}$ while measuring $g_{R}$ with $V_{L}=0$ fixed, and then sweeping $V_{L}$ while measuring $g_{L}$ with $V_{R}=0$ fixed.
After these sequential bias sweeps, $V_P$ was incremented and the process was repeated.
The resulting right-side tunneling conductance [Fig. 2(a)] exhibits a region of suppressed conductance at low bias, a characteristic of a superconductor, with discrete features inside the gap that oscillate as a function of plunger gate, a characteristic of Andreev bound states.
The left-side tunneling conductance [Fig. 2(b)] likewise exhibits a region of suppressed conductance at low bias, with discrete oscillating features inside the gap.
The subgap structure observed on the left and right sides are similar.
Directly comparing $g_{L}$ and $g_{R}$ at fixed gate voltage, as in Fig.~2(c), emphasizes that subgap peaks generally occur at the same bias voltage, suggesting that left and right peaks originate from the same Andreev bound state.
Although peaks are generally observed at the same bias voltages, the conductance associated with these peaks does not show a qualitatively clear correspondence.
On both the left and right side, the subgap conductance is not a symmetric function of bias, which is the focal point of a partner paper~\cite{menard_conductance-matrix_2019}.

It is interesting to note that the low-bias features in Figs.~2(a,b) are isolated from higher-lying states, stable at zero energy, and present on both ends of the device -- signatures traditionally associated with Majorana modes.
We emphasize that the data are taken at $B=0$, where topological effects are not generally expected \cite{lutchyn_majorana_2010,oreg_helical_2010}.

\begin{figure}
    \centering
    \includegraphics[width=0.48\textwidth]{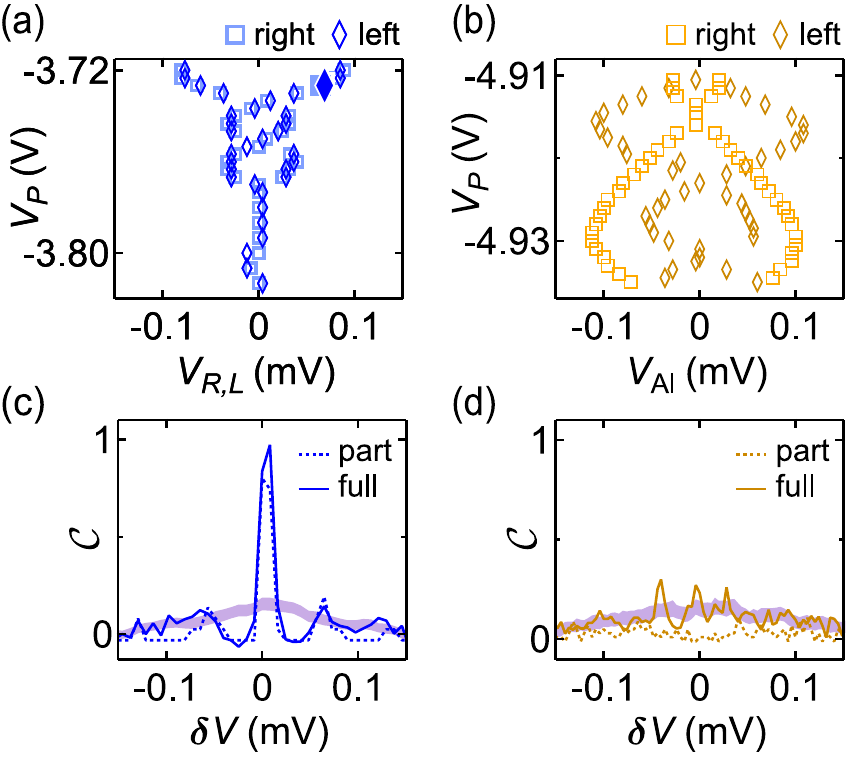}
    \caption{(a)  Extracted conductance-peak plunger voltage, $V_P$, and bias voltage, $V_{R,L}$ for the two lowest-energy states from short-device data in Fig.~2(a) [squares] and Fig.~2(b) [diamonds]. Solid square and diamond markers denote points indicated in Fig.~2(c).
    (b) Extracted conductance-peak plunger voltage, $V_P$, and bias voltage, $V_\mathrm{Al}$ for the two lowest-energy states from long-device data in Fig.~2(d) [squares] and Fig.~2(e) [diamonds]. Solid square and diamond markers denote points indicated in Fig.~2(f).
    (c) Correlator $\mathcal{C}$ of the binary peak masks for the data in Fig.~3a [dashed] and the full 2353-peak dataset [solid] from the short device.
    (d) Correlator $\mathcal{C}$ of the binary peak masks for the data in Fig.~3b [dashed] and the full 2058-peak dataset [solid] from the long device.
    Shaded regions represent $\pm 1 \sigma$ error bands derived from plunger-shifted data.}
    \label{fig3}
\end{figure}

\begin{figure}
    \centering
    \includegraphics[width=0.48\textwidth]{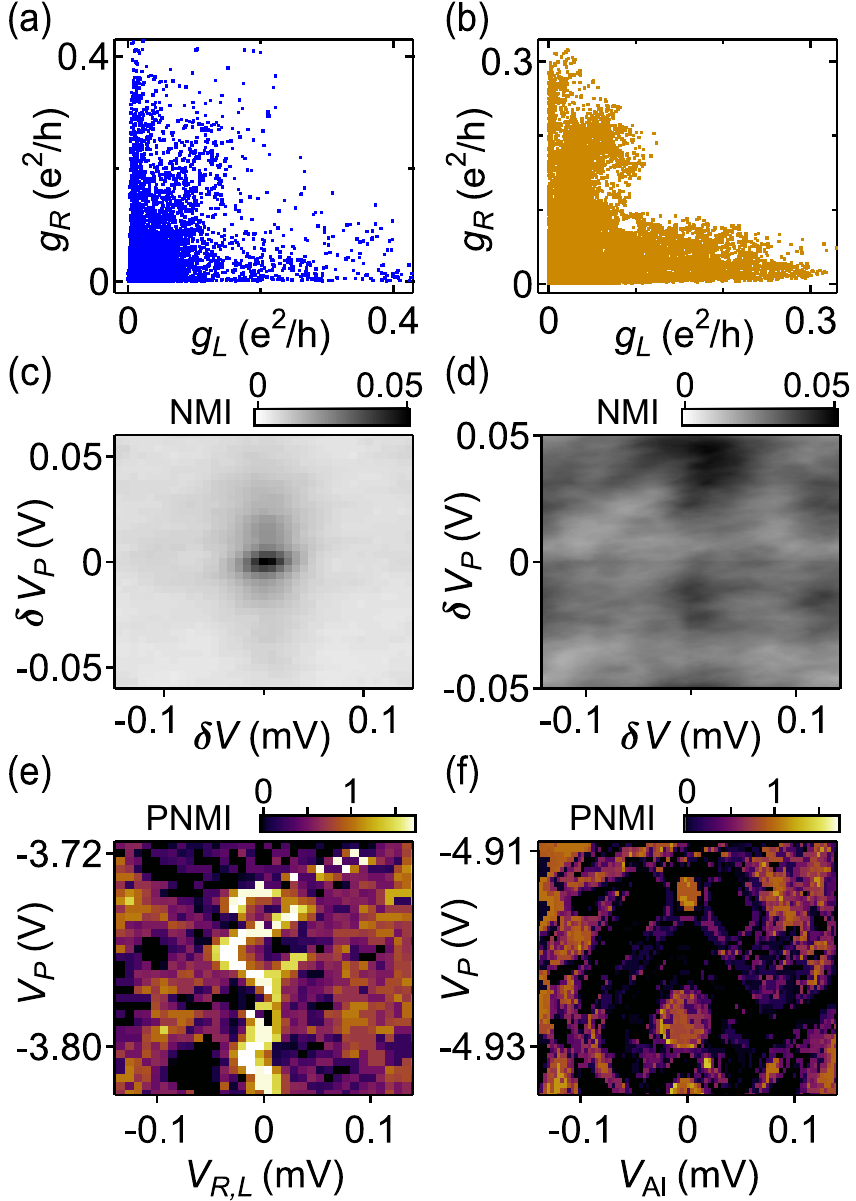}
    \caption{ 
    Sub-gap conductance pairs $(g_R, g_L)$ for short device bias $V_{R,L} < 0.15~\mathrm{mV}$ (a) and long device bias $V_\mathrm{Al} < 0.15~\mathrm{mV}$ (b).
    Normalized mutual information, $\mathrm{NMI}$, of the joint conductance distribution as a function bias-shift, $\delta V$, and plunger shift, $\delta V_P$ for the short device (c) and long device (d).
    Pointwise normalized mutual information, $\mathrm{PNMI}$, for short device subgap states (a) and for long device subgap states (b).}
    \label{fig4}
\end{figure}

Turning to the long device, $g_{R}$ versus gate and bias exhibits a gapped low-bias conductance with discrete subgap features [Fig.~2(d)], qualitatively similar to the short device.
The left-side conductance, shown in Fig.~2(e), also exhibits a gapped low-bias conductance and discrete subgap features.
Unlike the short device, in the long device the left and right subgap features exhibit markedly different gate-voltage dependences.
A direct comparison of $g_{L}$ and $g_{R}$ for the long device, shown in Fig.~2(f), confirms that subgap peaks on the left and right occur at different bias voltages, suggesting that the subgap features originate from different Andreev bound states.
We note that the long-device data are acquired differently than the short device data; by sweeping the aluminum bias $V_\mathrm{Al}$ and simultaneously measuring the currents on the left and right with $V_{R,L} = 0$.
This method has the advantage of simultaneous data acquisition on the left and right sides, but is disadvantageous in the short device, where the left and right biases must be carefully trimmed to zero to avoid complications from nonlocal effects.
We have directly compared the two methods on a small subset of the data and found that they are in agreement \footnote{see Supplement, incl Ref.'s \cite{Darbellay1999,Vinh2010,Kraskov2004,scikit-learn}}.
Taken together, the observations in Fig.~2 suggest a picture where individual subgap states can extend across the short device, but are associated with a single end of the long device.

To further analyze the conductance data, subgap peaks are identified using a peak finding algorithm over a wide range of gate voltages, of which the data in Fig.~2 is a small subset.
A total of 2353 peaks are extracted from the full short device dataset, and 2058 peaks are extracted from the full long device dataset.
A plot of the identified peak locations for each plunger-gate value from Fig.~2 demonstrates that the peaks are identified accurately, and emphasizes that the sugap peaks are indeed generally correlated for the short device [Fig.~3(a)], but are uncorrelated for the long device [Fig.~3(b)].

The correlations are quantified by introducing binary peak masks for the left side, $B_L( V, V_P )$, and the right side, $B_R( V, V_P)$ which take a value of 1 if a peak is identified at bias voltage $V$ and plunger voltage $V_P$ and a value of 0 otherwise.
The cross-covariance, $\mathcal{C}$, between the left and right binary masks is then computed, allowing for a bias offset, $\delta V$, between the two sides,
\begin{eqnarray}
\mathcal{C}( \delta V ) &=& \Big \langle B_L( V, V_P ) B_R( V + \delta V, V_P ) \Big \rangle \\*
\nonumber
    && - \Big \langle B_L( V, V_P ) \Big \rangle \Big \langle B_R( V, V_P ) \Big \rangle,
\end{eqnarray}
where $\langle \cdot \rangle$ denotes the expectation value with respect to both $V$ and $V_P$.
$\mathcal{C}$ serves as a correlation metric by counting the number of coincident peaks in each bias trace, averaged over plunger voltage.

In the short device, $\mathcal{C}$ is peaked around $\delta V=0$.
Including only the subset of data from Fig.~3(a) results in a correlator that is significantly less than unity [dashed trace, Fig.~3(c)], indicating that on average less than one peak is identified as being correlated at each gate voltage.
This is a result of the stringent definition of $\mathcal{C}$; it counts peaks as correlated only if they occur at identical bias voltages.
Indeed, inspection of the data in Fig.~3(a) reveals that every peak has a nearby partner, while only some occur at identical parameters.
Computing $\mathcal{C}$ for the full short-device dataset [solid trace, Fig.~3(c)] increases the average number of coincident subgap peaks to $\mathcal{C}=0.97$.
For large bias shifts, the correlator $\mathcal{C}$ has small fluctuations, consistent with the background level inferred by introducing a gate-voltage shift between the datasets.

In contrast, for the long device, no significant correlations are observed from the subset of dataset in Fig.~3(b) [dashed trace, Fig.~3(d)], and a small correlation peak of $\mathcal{C}=0.27$ is resolved when averaging over the full dataset [solid trace, Fig.~3(d)].
The observed drop in correlations, by a factor of $3.7$, presumably reflects a characteristic length-scale for subgap states in these structures.
By examining subsets of the long device dataset, we have identified a small region in gate voltage that gives rise to the correlations \cite{Note1}, demonstrating that $\mathcal{C}$ is a useful tool for identifying rare features in the data.

While the presence of subgap features in the short device is strongly correlated between both ends, the conductance associated with these features fluctuates strongly.
To study correlations in these fluctuations, each measurement of the right and left conductances is visualized as a point in the $(g_R, g_L)$ plane.
The subgap $(g_R, g_L)$ data for the short device [Fig.~4(a)] is indeed widely distributed, with a long tail extending out to $g_{R,L} \sim 0.4~\mathrm{e^2/h}$. 
In the long device, the subgap $(g_R, g_L)$ data are also broadly scattered, but the $(g_R, g_L)$ pairs tend more towards the axes in the $g_R,g_L$ plane [Fig.~4(b)].
In other words, there are more points with large weight on both sides in the short device as compared to the long device, which gives a preliminary indication of positive correlations in conductance for the short device.

To quantify this observation, joint distributions, $P( g_R, g_L )$, and marginal distributions, $P(g_R)$ and $P( g_L )$, are estimated from the scatter-plot data.
In general, for independent variables one expects $P( a, b )=P(a) P(b)$.
Deviations from this relationship are quantified by the normalized mutual information, $\mathrm{NMI}$, which, in analogy with a correlator, is calculated as a function of bias and plunger shift between the left and right datasets, $\mathrm{NMI}=\mathrm{NMI}( \delta V, \delta V_P)$ \cite{Note1}.

For the short device, the mutual information is strongly peaked at $\delta V = \delta V_P = 0$ [Fig.~4(c)], indicating the presence of correlations in the conductance distributions.
As a function of bias shift, $\delta V$, the mutual information decays with a $1/e$ characteristic length that approximately matches the observed width of subgap peaks, suggesting that the conductance correlations are associated with subgap states.
As a function of plunger shift, $\delta V_P$, the mutual information also decays sharply, again consistent with contributions from subgap states.
In contrast, for the long device, the mutual information is relatively flat and featureless [Fig.~4(d)], suggesting that there are no conductance correlations originating from the subgap states.
The mutual information for the long device, however, still differs significantly from zero, indicating there are broad, gate and bias-voltage independent features that are mutually dependent between the two ends. 

A more granular view of the conductance correlations is obtained by visualizing the pointwise normalized mutual information, $\mathrm{PNMI}$, which is defined by the relation $\mathrm{NMI} = \overline{ \mathrm{PNMI} }$, where the overline denotes an expectation value over $P( g_R, g_L )$.
Whereas $\mathrm{NMI}$ is a property of the entire distribution, $\mathrm{PNMI}$ is associated with individual $( g_R, g_L )$ pairs, and can therefore be mapped as a function of the measurement parameters.
In the short device, a map of the $\mathrm{PNMI}$ as a function of gate and bias voltage reveals regions of elevated $\mathrm{PNMI}$ corresponding to the stable low-lying bound state identified in previous discussion [Fig.~4(e)].
This bound state also has an elevated $\mathrm{PNMI}$ compared to other correlated features throughout the entire dataset \cite{Note1}, suggesting that although many features in the short device are measurable from both device ends, additional correlations in the conductance of these features are relatively rare.
In contrast to the short device, for the long device the $\mathrm{PNMI}$ is smaller, although there is still some weak structure [Fig.~4(f)]. 
The conductance correlations are not thoroughly understood, although, qualitatively, correlations in the conductance could be expected as a result of states that span the entire device.

In summary, we interpret the well-correlated subgap states in the short device as a clear indication that bound states extend several hundred nanometers into the hybrid region, essentially spanning the entire device.
If the zero-field bound states are attributed to an accidental potential well in the tunneling region \cite{deng_majorana_2016}, then their spatial extent would indicate a soft confinement, which will strongly effect the behaviour at nonzero magnetic field \cite{kells_near-zero-energy_2012,prada_transport_2012,chun-xiao_andreev_2017}.
Following this reasoning, we anticipate that these measurements could be used to refine the theoretical understanding of topologically trivial subgap states, including in numerical simulations, and possibly aid in distinguishing them from the topological case.

Looking ahead, the techniques developed here can be applied in the putative topological regime.
Computing a correlator after feature extraction, as in Fig.~(3), effectively removes the background from conductance fluctuations.
Quantifying conductance-correlations using the mutual information requires minimal analysis of the data, and explicitly tests for correlations in the conductance values of all data points.

\begin{acknowledgments}
We thank Karsten Flensberg, Martin Leijnse, Dmitry Pikulin, Roman Lutchyn, and Ga\v{s}per Tka\v{c}ik for valuable discussions.
Research was supported by Microsoft Corporation and the Danish National Research Foundation.
\end{acknowledgments}

\clearpage
\newpage

\begin{figure}
\includegraphics[page=1,trim={0.7in 0.7in 0.7in 0.7in},width=1.0\textwidth]{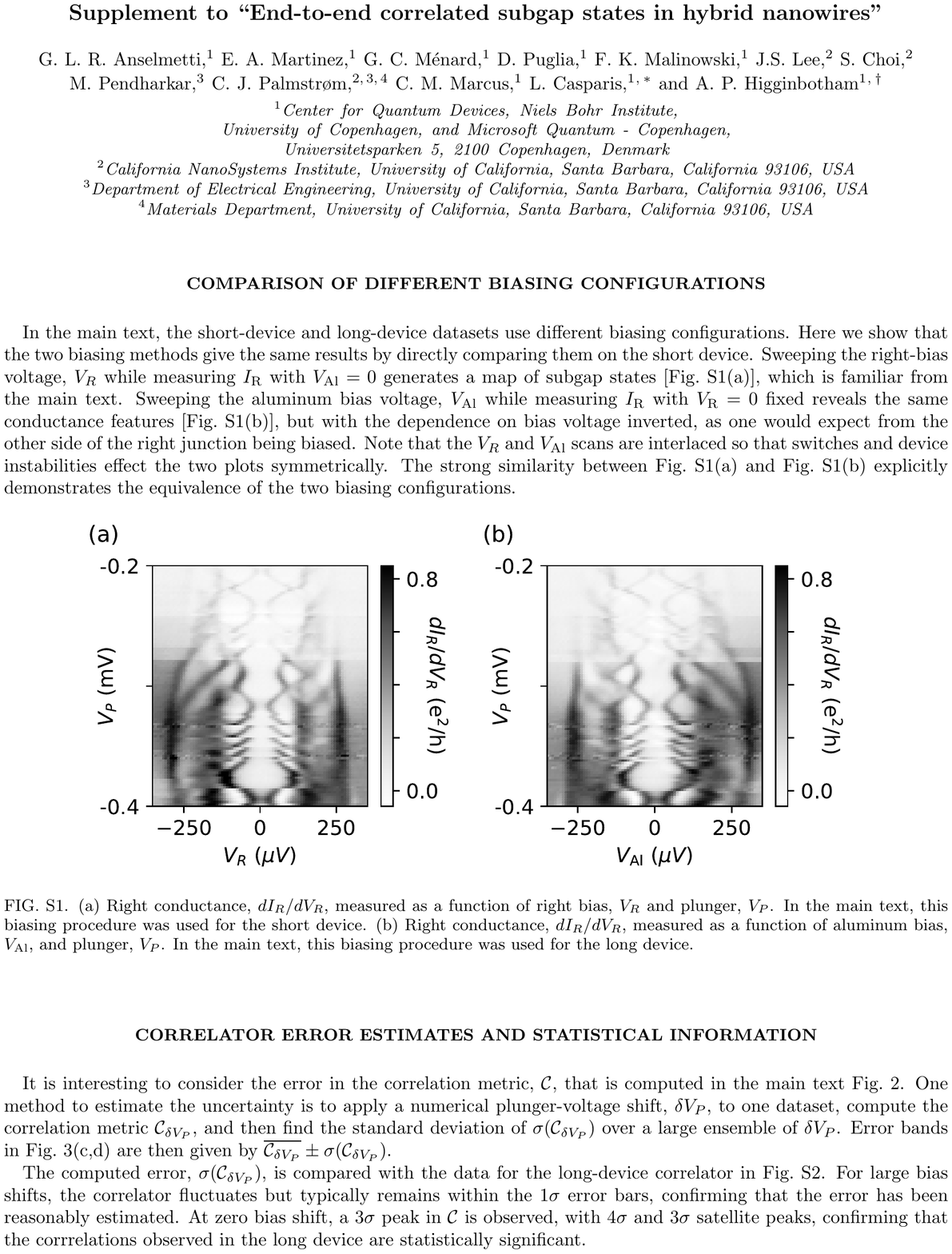}
\end{figure}
\begin{figure}
\includegraphics[page=2,trim={0.7in 0.7in 0.7in 0.7in},width=1.0\textwidth]{supplement}
\end{figure}
\begin{figure}
\includegraphics[page=3,trim={0.7in 0.7in 0.7in 0.7in},width=1.0\textwidth]{supplement}
\end{figure}
\begin{figure}
\includegraphics[page=4,trim={0.7in 0.7in 0.7in 0.7in},width=1.0\textwidth]{supplement}
\end{figure}

\end{document}